%% file: sample-authordraft.tex
\documentclass[sigconf]{acmart}

\usepackage{enumitem}
\usepackage{booktabs}
\usepackage{tabularx}
\usepackage{xcolor,colortbl}
\usepackage{tabularx}
\definecolor{lightgray}{gray}{0.93}
\definecolor{slightgray}{gray}{0.98}
\definecolor{darkgray}{gray}{0.77}
\usepackage{makecell}
\usepackage{multirow}
\usepackage{threeparttable}
\usepackage{hyperref}
\usepackage{graphicx}
\RequirePackage[normalem]{ulem} 
\RequirePackage{color}\definecolor{RED}{rgb}{1,0,0}\definecolor{BLUE}{rgb}{0,0,1} 

\newcolumntype{Y}{>{\centering\arraybackslash}X}
\AtBeginDocument{%
  \providecommand\BibTeX{{%
    \normalfont B\kern-0.5em{\scshape i\kern-0.25em b}\kern-0.8em\TeX}}}

\begin{document}

\title{MindScratch: A Visual Programming Support Tool for Classroom Learning Based on Multimodal Generative AI}
\author{Yunnong Chen}
\affiliation{%
  \institution{College of Computer Science and Technology, Zhejiang University}
  \city{HangZhou}
  \state{ZheJiang}
  \country{China}
}

\author{Shuhong Xiao}
\affiliation{%
  \institution{College of Computer Science and Technology, Zhejiang University}
  \city{HangZhou}
  \state{ZheJiang}
  \country{China}
}

\author{Yaxuan Song}
\affiliation{%
  \institution{College of Computer Science and Technology, Zhejiang University}
  \city{HangZhou}
  \state{ZheJiang}
  \country{China}
}

\author{Zejian Li}
\affiliation{%
  \institution{School of Software Technology, Zhejiang University}
  \city{HangZhou}
  \state{ZheJiang}
  \country{China}
}
\author{Lingyun Sun}
\affiliation{%
  \institution{College of Computer Science and Technology, Zhejiang University}
  \city{HangZhou}
  \state{ZheJiang}
  \country{China}
}

\author{Liuqing Chen\textsuperscript{*}}
\affiliation{%
  \institution{College of Computer Science and Technology, Zhejiang University}
  \city{HangZhou}
  \state{ZheJiang}
  \country{China}
}
\email{chenlq@zju.edu.cn}
\renewcommand{\shortauthors}{}

\begin{abstract}
Programming has become an essential component of K-12 education and serves as a pathway for developing computational thinking skills. Given the complexity of programming and the advanced skills it requires, previous research has introduced user-friendly tools to support young learners. However, our interviews with six programming educators revealed that current tools often fail to reflect classroom learning objectives, offer flexible, high-quality guidance, and foster student creativity. This highlights the need for more adaptive and reflective tools. Therefore, we introduced MindScratch, a multimodal generative AI (GAI) powered visual programming support tool. MindScratch aims to balance structured classroom activities with free programming creation, supporting students in completing creative programming projects based on teacher-set learning objectives while also providing programming scaffolding. Our user study results indicate that, compared to the baseline, MindScratch more effectively helps students achieve high-quality projects aligned with learning objectives. It also enhances students' computational thinking skills and creative thinking. Overall, we believe that GAI-driven educational tools like MindScratch offer students a focused and engaging learning experience.

\end{abstract}





\maketitle
\input{sections/introduction}
\input{sections/related_work}
\input{sections/formative_study}
\input{sections/system}

\input{sections/exp}

\input{sections/results}
\input{sections/disccusion}
\input{sections/conclusion}

\section*{Acknowledgements}
We express our sincere gratitude to Gang Dai, Wen Meng, and Xuenan Jiang from AC Source Programming for their expert educational perspectives and assistance with the experiment and; to all the user study participants for their time and contribution. 
\section*{Funding}
This research was funded by the National Natural Science Foundation of China (Grant No. 62207023), and the Ng Teng Fong Charitable Foundation in the form of ZJU-SUTD IDEA (Grant No. 188170-11102).
\section*{Disclosure statement}
No potential conflict of interest was reported by the author(s).

\bibliographystyle{ACM-Reference-Format}
\bibliography{sections/Reference}

\clearpage
\appendix
\section{Appendix A}
\begin{table}[h]
    \centering
    \caption{Detailed descriptions of the two theme-based creative programming tasks.}
    \begin{tabular}{|c|m{4cm}|m{4cm}|m{8cm}|}
        \hline
        N & Project Description & Learning Objectives & The Scratch Code and Creations\\
        \hline
        1 & Cat and Mouse: Create a maze background with two characters, a cat and a mouse. When the cat touches the mouse, the cat grows larger. Both the cat and the mouse can eat cheese, and the mouse will randomly appear at any location in the maze. & 
        (1) Students can proficiently use various methods to control the cat or mouse. \newline
        (2) Students can use relevant code for detection and decision-making functions to complete the game's features. & 
        \includegraphics[width=0.4\textwidth]{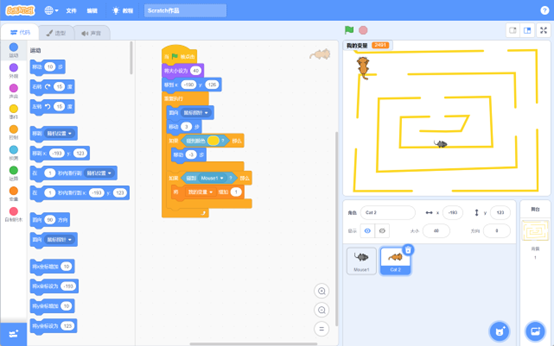} 
        \includegraphics[width=0.4\textwidth]{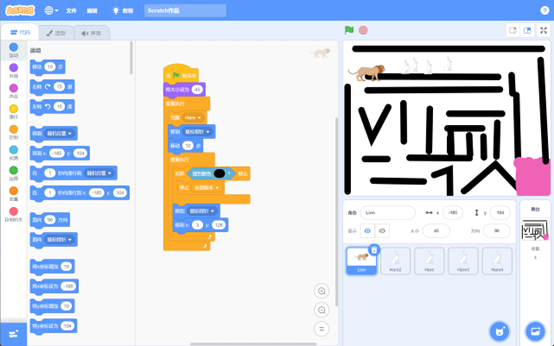}\\
        \hline
        2 & Kitten Fishing: Create a game where a kitten is fishing in a pond. There are four characters: the kitten, the fishing rod, the fish, and the score. When the fishing rod catches a fish, the score increases by one. & 
        (1) First, set up the pond stage. \newline
        (2) Second, make the fish appear anywhere in the pond and enable them to move back and forth in the water.  \newline
        (3) Third, control the fishing rod in the kitten’s hand to move downwards. If the rod touches a fish, the fish should disappear, and the score counter should increase by one. & 
        \includegraphics[width=0.4\textwidth]{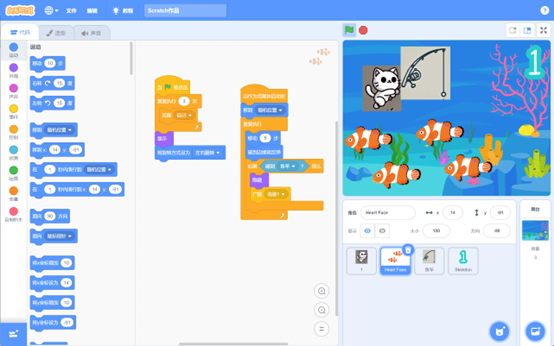}\\
        \hline
    \end{tabular}
    \label{tab:my_label}
\end{table}

\end{document}

%% file: sections/introduction.tex
\section{Introduction}


In the past two decades, computational thinking (CT) has become a focal point in educational research and practice \cite{SHUTE2017142, hsu2019computational}. While CT and programming are distinct concepts, most educational practices regard learning programming as an effective way to cultivate CT skills \cite{voogt2015computational}. This perspective underscores the importance of early exposure to CT in shaping children's futures, leading to a recent trend of teaching programming to young learners \cite{lonka2018phenomenal}. Accompanying this development is the emergence of programming languages designed specifically for young learners, with Scratch being a prime example \cite{resnick2009scratch}. Scratch is widely used for creative programming, with projects often involving games, stories, and animations \cite{romero2017computational}. However, educators face a trade-off between rigid knowledge transmission and engaging project-based learning when teaching programming. Effectively guiding learners to stay aligned with predefined learning objectives and providing personalized feedback during project-based learning presents a significant challenge.

Previous research has explored methods of using advanced programming learning tools to support children's creative programming in self-directed learning. These tools can be categorized into programming concept learning tools and block-based programming support tools. For instance, Visual StoryCoder \cite{dietz2023visual} focuses on mastering abstract concepts in visual programming. CodeToon \cite{suh2022codetoon} is dedicated to transforming abstract code into engaging stories and comics.
ChatScratch \cite{chen2024chatscratch} supports creative programming by providing both creative stimulation and code guidance. However, without structured courses and instructional guidance, the potential for meaningful learning outcomes is significantly diminished \cite{10.1145/1592761.1592779,10.1145/1999747.1999796,popovici2023chatgpt}.

To understand the specific needs of creative programming in the classroom and assess the effectiveness of current programming support tools in such contexts, we conducted a formative study with six programming educators. Educators evaluated the applicability of these tools in classroom settings from the perspective of structured course design, focusing on objectives, processes, and content \cite{robins2003learning}. The formative study revealed three key insights: (1) Strategies teachers use to ensure students achieve their learning goals during project-based learning. (2) Approaches for providing customized creative and programming support. (3) Methods to make creative resources more accessible to students while remaining manageable for teachers. Additionally, we found that previous programming support tools lacked controllable guidance for students, hindering alignment with course learning objectives and delaying timely feedback from teachers. This issue may result in teachers spending excessive time addressing simple questions instead of offering constructive feedback promptly.

\begin{table*}[bht]
\centering
\caption{Comparison of MindScratch with multiple programming learning tools based on educators' feedback, with a focus on their applicability in classroom programming instruction.}
\vspace{-0.11in}
\begin{tabular}{l|c|c|c|c}
\hline
\multicolumn{5}{c}{} \\[-8pt]
\textbf{Targets} & \textbf{CodeToon \cite{suh2022codetoon}} & \textbf{Visual StoryCoder \cite{dietz2023visual}} & \textbf{ChatScratch \cite{chen2024chatscratch}} & \textbf{MindScratch} \\
\multicolumn{5}{c}{} \\[-8pt]
\hline
\multicolumn{5}{c}{} \\[-8pt]
Reflect Class Objectives & \checkmark & $\times$ & $\times$ & \checkmark \\
Task Breakdown and Progression & $\times$ & \checkmark & $\times$ & \checkmark \\
Access to Multimedia Materials & \checkmark & $\times$ & \checkmark & \checkmark \\
Controlled Learning Guidance & $\times$ & $\times$ & $\times$ & \checkmark \\
Support Teachers to Provide Timely Feedback & $\times$ & $\times$ & $\times$ & \checkmark \\
\multicolumn{5}{c}{} \\[-8pt]
\hline
\end{tabular}
\vspace{-0.11in}
\label{tab:comparison}
\end{table*}

Based on these insights, this paper introduces MindScratch, a programming learning tool designed to support students in creating personalized, high-quality programming projects in classroom settings. Table \ref{tab:comparison} summarized the applicability of MindScratch and other programming learning tools in classroom settings across five key aspects: alignment with class objectives, task breakdown and progression, access to multimedia materials, controlled learning guidance, and support for teachers in providing timely feedback. To better help students achieve their learning objectives, teachers can set initial task themes and learning objectives in MindScratch. The tool receives these inputs and presents them in the form of a mind map. Leveraging the system prompt capabilities of large language models (LLMs), MindScratch retains the teacher-defined learning objectives throughout student interactions. Guided by an LLM-driven conversational agent, MindScratch provides students with controlled learning guidance rather than simply answering all of their questions.

To inspire students' creativity, MindScratch encourages the generation of multimodal creative materials—such as text, images, audio, and code—through conversations aligned with the classroom workflow, covering project planning, material creation, and code implementation. These materials are displayed as a mind map based on the classroom workflow structure. MindScratch helps students break down teacher-set learning goals into sub-tasks, manage the learning process, use generated materials for personalized visual effects, and complete programming projects with code suggestions. Notably, with MindScratch's support, teachers can shift their focus from managing numerous students' project progress to providing timely and constructive feedback. To develop an effective Scratch educational LLM, we collaborated with educators to create a specialized dataset and fine-tuned the GPT-3.5-turbo model to provide scaffolded guidance. Additionally, MindScratch employs a text-to-image generative model \cite{StableDiffusion2023} and a text-to-audio generative model \cite{ghosal2023tango} to help students access high-quality creative materials.

To evaluate MindScratch's effectiveness in facilitating visual programming in the classroom, we conducted a within-subject study involving 24 fifth-grade students. The results indicated that, by using MindScratch, students more effectively achieved their learning objectives, showing significant improvements in CT skills and creativity. The extensive expansion of mind map nodes demonstrated deep engagement and valuable explorations by the students. Additionally, long-term monitoring of three students, combined with semi-structured interviews with six programming educators, provided insights into the future adoption and integration of AI assistants like MindScratch into AI-supported curricula.

In summary, the main contributions of this work are:

\begin{itemize} 

\item Through a formative study, we identified the challenges of aligning objectives, managing processes, and delivering content in creative classrooms, along with the limitations of current programming support tools.

\item We introduced MindScratch \footnote{\url{https://github.com/ArthurWish/MindScratch}}, a visual programming support tool that employs interactive mind maps powered by multimodal generative AI. This approach provides interactive creativity exploration, scaffolded code assistance, and the generation of creative materials, thereby enhancing students' learning outcomes in the creative classroom.

\item We conducted a comparative study between MindScratch and Scratch, highlighting MindScratch's effectiveness in enhancing students' programming learning. We also provide insights and design considerations for future creative programming support tools. \end{itemize}

%% file: sections/related_work.tex
\section{Related Work}

\subsection{Generative AI in Computing Education}

Generative artificial intelligence (GAI) has become a significant topic of debate, particularly due to its ability to produce high-quality artistic media across visual arts, concept art, music, literature, video, and animation. For instance, diffusion models can synthesize high-quality images \cite{StableDiffusion2023}, and large language models (LLMs) can generate coherent and impressive prose and poetry in various contexts \cite{brown2020language}.

In the context of computing education, GAI offers several advantages. One major benefit is the enrichment of learning resources. Sarsa et al. \cite{sarsa2022automatic} demonstrated the effectiveness of OpenAI Codex in generating personalized code exercises and complex code explanations, thereby reducing teachers' preparatory workload. MacNeil et al. \cite{macneil2022generating} further explored how GAI can provide code explanations by examining the relationship between different prompts and the resulting explanations. Another advantage lies in supporting the programming process, notably through the provision of exemplar solutions. These solutions offer students an easily understandable starting point \cite{kazemitabaar2023studying,chen2024chatscratch}, enhancing their task performance. Additionally, students can learn from these examples by observing various problem-solving approaches, coding styles, and program designs \cite{finnie2022robots,indriasari2020review}.

However, it is crucial to address the challenges and negative impacts of GAI in educational practice. Researchers have expressed considerable concern about the misuse of GAI, especially among novices introduced to programming in the era of such technology \cite{denny2024computing,kasneci2023chatgpt,yan2024practical}. Chen et al. \cite{chen2021evaluating} highlighted potential harmful biases in generated code, including issues like racism and stereotypes. Security concerns are another significant issue. Moreover, instances of academic dishonesty have increased, with students turning to GAI to complete coding assignments \cite{finnie2023my,finnie2022robots}. The variability of GAI-generated code makes such cheating more difficult to detect \cite{simon2016negotiating}. Over-reliance on GAI may also reduce students' critical thinking, as novices might adopt generated solutions without thoroughly engaging with the material, which is detrimental to learning processes based on reinforcement \cite{becker2023programming}. To mitigate these issues, we employ mind maps to visualize the project development process, collaboratively developing the mind map and programming project with students in a step-by-step approach.

\subsection{Learning Computational Thinking through Programming}

To meet the growing demand for 21st-century skills, efforts are being made worldwide to integrate computational thinking (CT) into school education \cite{hsu2019computational,tikva2021mapping}. While programming is not the only way to teach CT, its practical methods and strong connection to CT concepts have made it a key approach in classrooms \cite{voogt2015computational}. Some researchers argue that programming should be seen as a teaching strategy rather than just a technical skill or set of coding techniques. Based on this view, integrating programming with classroom instruction can lead to more meaningful outcomes \cite{romero2016move,meerbaum2011habits}.

Efforts are also being made to reduce teaching workloads by using AI systems to handle basic tasks. For example, VizProg \cite{zhang2023vizprog} offers a system that helps teachers monitor students' coding progress in real time, allowing them to assist students who may be struggling. Tools like PuzzleMe \cite{wang2021puzzleme} and VizPI \cite{tang2023vizpi} promote peer assessment in programming classes, encouraging collaboration and code evaluation among students. Additionally, many other tools guide students in areas such as variable declaration \cite{glassman2015foobaz}, coding style \cite{moghadam2015autostyle}, and code correctness \cite{singh2013automated}.

To support the widespread adoption of programming education in K-12, particularly in the early grades, significant efforts have been made. Due to challenges related to literacy demands and code conventions, block-based programming \cite{reilly2013kindergarten,resnick2009scratch,klassner2003lego} has become a popular method for teaching CT. These approaches are often combined with specific tasks like robot programming \cite{gordon2015designing,rivera2023toward} or creative programming \cite{chen2024chatscratch,romero2017computational} to meet learning goals and increase student engagement. To address the lack of collaboration in programming systems, Wang et al. \cite{wang2024lighting} introduced a new tangible, collaborative programming system called Lighters. However, classroom teaching is a structured process where teachers carefully plan lessons for different periods, and previous studies have not adequately provided controlled learning guidance to support teaching. This study explores the use of mind maps as a visual tool to help teachers better manage students’ progress and assist students in organizing project ideas and programming materials more effectively.

\subsection{Assisting Novices in Learning Programming}

Romero et al. \cite{romero2017computational} divided programming learning into five stages based on learner engagement. The first stage involves passive exposure to teacher-led explanations, videos, or tutorials, where learners are not yet involved in the creative process. This stage is supported by widely available open-source videos and tutorials. The second stage introduces step-by-step programming activities, allowing learners to follow instructions to achieve predefined goals. Platforms like Code.org \cite{code.org} and Hour of Code \cite{hourofcode} provide numerous projects for novices to gain hands-on experience. Recently, LLMs have also been used to generate tutorials for simple tasks \cite{qi2023conversational,kazemitabaar2023novices}.

The third stage involves creative coding, where learners use programming tools to develop original works, integrating knowledge across multiple domains. Tools at this level typically offer creativity support and process control to balance task completion with divergent thinking \cite{dietz2023visual,chen2024chatscratch,dietz2021storycoder,wang2023colaroid}. Building on this framework, our study investigates the potential of conversational agents powered by generative AI to provide structured guidance aligned with classroom workflows while delivering an engaging creative programming learning experience.

%% file: sections/formative_study.tex
\section{Formative Study}
Our formative study aims to identify the challenges teachers and students face in creative programming classrooms, as the teaching is project-based, support-consuming, and content resource-consuming, which is special compared to other classroom settings.

\subsection{Procedure}
We recruited six educators (3 females, and 3 males) of age 26-39 (M=31, SD=4.44), each with over four years of experience in teaching programming to students and proficient in multiple programming languages, including Scratch, Python, and C. All educators are affiliated with a prominent programming education institution in China and have substantial experience in leading both small after-school courses (3-9 students) and larger school-based courses (over 20 students).

We conducted one-on-one discussions with each educator, introducing them to three representative programming support tools: CodeToon \cite{suh2022codetoon}, Visual StoryCoder \cite{dietz2023visual}, and ChatScratch \cite{chen2024chatscratch}, all of which were presented at prominent conferences (e.g., CHI and UIST) in the past two years. CodeToon allows teachers to bridge the gap between abstract coding concepts and tangible learning outcomes by transforming symbolic text code into engaging stories and comics. This helps students visualize and better understand complex ideas. Visual StoryCoder offers a unique, block-based language that simplifies programming concepts, ideal for teachers aiming to introduce foundational programming skills. ChatScratch, utilizing generative artificial intelligence, enables students to actively engage in creating story-based projects in Scratch, fostering creativity and problem-solving skills in a supportive, AI-enhanced environment.

Before our discussions, none of the educators had any knowledge of these three applications. To further prevent the researchers' biases from influencing their attitudes, we facilitated their understanding of these applications in two ways. First, we presented video materials of these works, which were uploaded to the ACM Digital Library to the educators. Second, the educators themselves experienced the usage of these tools. For tools that were not open-source, we conducted simulation using Wizard of Oz (WOz) techniques \cite{garcia2023support}. The simulations were based on the descriptions provided in the methods section of the respective papers and application usage procedures (if exist) documented in their supplementary materials.

\subsection{Findings}

Systematic programming teaching is typically anchored in a curriculum that is meticulously designed by teachers, encompassing the objectives, process, and content of each lesson. In this case, educators expect supporting tools to enhance student's learning experience and provide additional convenience while being adaptable to the existing teaching approach. As shown in Table \ref{tab:comparison}, a summarized takeaway for our study is that these three representative tools faced challenges to be directly applied within classroom settings for creative programming. Based on feedback from educators, we report the main issues as below.

\subsubsection{Objectives.}

Educators expect the support tools to align with the specific objectives they have designed, rather than allowing students to engage in unrestricted activities. This necessitates that teachers possess the ability to directly tailor and intervene with the support tools. For example, as P2 suggested, teachers should be able to \textit{``specify the knowledge points that need to be addressed or the themes of programming tasks in the current lesson.''} In this context, CodeToon stands out for its ability to effectively facilitate the exploration and understanding of designated code fragments. Conversely, Visual StoryCoder does not offer the functionality for educators to pre-define learning objectives, which means its pedagogical effectiveness is wholly dependent on the learner's self-directed engagement. Regarding ChatScratch, it accommodates the specification of programming themes, yet due to its focus on supporting creativity with AI-generated content, the control over programming project workflows is notably compromised. 

\subsubsection{Process.}
Classroom time is always limited, and teachers must wisely allocate learning time for each key objective, striking a balance between fostering students' divergent exploration and advancing the process. In this case, educators prioritize the system's ability to function as a personal supervisor for each student, managing the learning pace and keeping students focused. For instance, \textit{``the system could provide a detailed breakdown of steps based on the task objectives and guide students through completing each item sequentially''}, as noted by P1. This ensures that students progress through their lessons efficiently, adhering to a structured timeframe, and keeps the class on track. In this regard, Visual StoryCoder offers process guidance through a conversational agent, clarifying the objectives of each step, thereby facilitating task progression. Conversely, CodeToon and ChatScratch allow for free exploration, which may lead students to focus on refining comic details or generating assets, diverting attention from their primary goal of programming. 

Educators also highlighted key barriers that slow the class process, including aspects related to creativity, such as the conception of project roles and plots, as well as parts related to coding, such as the implementation of logic and data flow. Each student faces personalized issues, requiring teachers to address them specifically. \textit{``Only after resolving everyone's confusion, the class can progress to the next step.''}, as said by P1. In this case, they expected the system to provide support when critical confusion arose, pulling students out of any standstill. For CodeToon and Visual StoryCoder, the answer is negative. ChatScratch provides two-stage assistance in its code support process, including suggestions for code selection and code implementation. Nonetheless, educators believe that the generated code in this process remains difficult to comprehend, especially for beginners.

\subsubsection{Content}

In the field of programming education, project-based learning (PBL) has become a highly popular teaching method. When students are learning by solving real-world problems, they always need to utilize additional content resources. For instance, when constructing a simple website, students may require additional textual and decoration materials; while creating animations in Scratch could involve the introduction of character images and audio materials. Generally, these materials are prepared in advance by teachers, although students are sometimes dissatisfied with them due to a lack of choice and personalization. While allowing students to source their materials online can be too time-consuming and may lead to classroom disorder. Within this context, educators expect support tools to facilitate access to the additional materials needed for the projects. On this matter, CodeToon is capable of providing additional stories and comics, yet there is a convergence issue in terms of form; While Visual StoryCoder lacks this feature. Leveraging the artistic capabilities of generative AI, ChatScratch can assist students in obtaining the necessary materials, with its support primarily focused on images.

The quality of materials provided by the system is also a major concern for educators. They need to be accurate, non-toxic, and appropriate for students. In this regard, educators believe that CodeToon can meet their needs because it relies on established rules to parse code and utilizes confirmed materials to create stories and comics. Regarding Visual StoryCoder, educators express concerns due to its reliance on students' active storytelling and drawing. Without proper guidance, there is a risk that students might create or come across content that includes inappropriate language, imagery, or concepts. For ChatScratch, researchers suggest the need for additional mechanisms to review AI-generated content.

\subsection{Design Goals}
Building upon the insights from our investigations, we derived three principal design goals for our system to support students in creative programming learning in the classroom: 

\begin{itemize}
\item \textbf{DG1: Support alignment between the project development process and learning objectives:} 
The system should ensure that students' project creation process stays aligned with the classroom learning objectives set by the teacher. Therefore, the system needs to display the teacher's requirements, such as the story elements and code blocks that must be included in the project. To ensure that students pay attention to and understand these key elements and code blocks, the system should highlight this information and provide necessary explanations.

\item \textbf{DG2: Provide guidance and real-time support for programming learning at scale:} To reduce the pressure on teachers in maintaining students' progress alignment and allow them to focus more on providing constructive feedback, the system should offer project development support for students. For example, it could visualize the students' ideation process, guide them in breaking down tasks, and offer suggestions for code implementation.

\item \textbf{DG3: Provide creative programming support based on classroom projects:} In creative programming classes, students can build upon project templates provided by the teacher. However, the limited programming resources offered by the teacher may not meet the diverse creative needs of students, and teachers may lack the time to help each student acquire custom images or audio materials. Therefore, the system should provide multimodal creative material support, leveraging the generative capabilities of GAI to help students realize personalized programming projects.
\end{itemize}

%% file: sections/system.tex
\section{MindScratch}

Based on the design objectives, we developed MindScratch, a GAI-driven visual programming learning system to help students achieve creative programming projects aligned with classroom goals. Our system uses a mind map to display the project objectives set by the teacher. Teachers can create an initial mind map in the system and input the learning objectives through an input box. For example, a teacher can set a theme like "Kitten Fishing" (red node, Fig. \ref{fig:UI}.(a.3)) and course objectives (such as learning conditions and loops, gray nodes, Fig. \ref{fig:UI}.(a.3)) within the mind map. The LLM in the system receives the learning objectives as a system prompt, which serves as the context for generating subsequent content. Through the LLM-powered conversational agent (Fig. \ref{fig:UI}.b, chat box), the system guides students in task breakdown, sparks creativity, and offers coding implementation suggestions based on the learning goals set by the teacher. The content created by students during their interaction with the agent is then presented in the mind map (Fig. \ref{fig:UI}.a, mind map). Finally, students complete creative programming tasks based on the mind map they created. For example, they can import character images from the mind map into the Scratch software, use generated audio to enhance the storytelling aspect of their programming, and implement program functionalities according to the programming logic and code suggestions on the mind map. The advantage of the mind map is that it reduces students' cognitive load by breaking down the process of simultaneously conceptualizing ideas and programming into two distinct steps: first brainstorming in the mind map, and then proceeding to program.

\subsection{MindScratch's User Interface}

\begin{figure*}[thp]
    \centering
\includegraphics[width=0.9\textwidth]{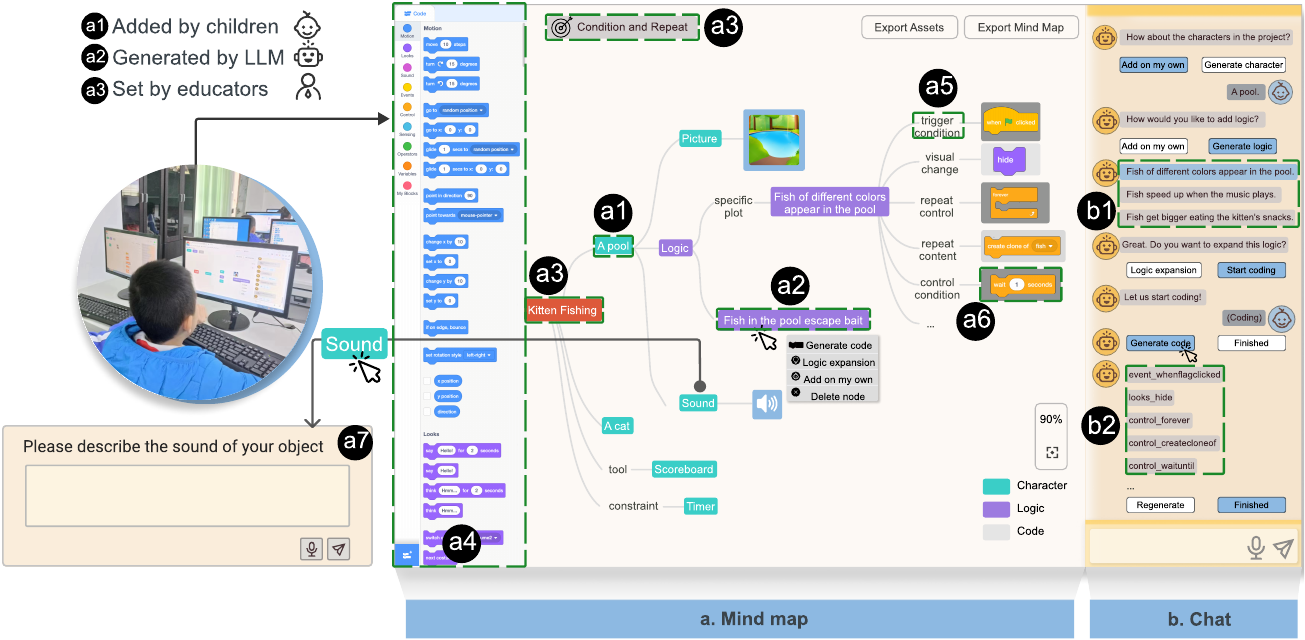}
    \vspace{-0.11in}
    \caption{\textbf{MindScratch's User Interface. In the mind map (a) with block palette (a4), each node represents the collaborative creation of students, teachers, and AI (a1, a2, a3). The connections between nodes are annotated, and the highlighted nodes indicate their relevance to learning objectives (a5, a6). Within the dialogue box (b), the system provides structured guidance and real-time support through a Q\&A mechanism (b1, b2). A floating text input box allows for the generation of audio materials through text editing (a7).}}
    \vspace{-0.11in}
    \label{fig:UI}
\end{figure*}

Figure \ref{fig:UI} illustrates the user interface of MindScratch, which is composed of three primary panels: an interactive mind map with a block palette, a chat box, and a drawing board (Figure \ref{fig:UI2}). The mind map allows nodes to be added in two ways. Users can either right-click on a selected node to choose a generated node type (character, logic, or code), or they can follow the suggestions provided by the agent during conversations. This visual representation helps students organize their projects. It also aids teachers in monitoring individual progress and aligning learning objectives. To create an engaging learning experience, we integrated a chat box where the agent provides guidance and suggestions to help students construct their projects through interactive conversations. Additionally, we developed a drawing board and a text-to-audio generation feature, offering students easy access to high-quality programming assets. 

\subsubsection{Mind Map view}

Existing research indicates that mind maps are effective tools for learners to grasp complex concepts, fostering engagement in critical, higher-order, and reflective thinking \cite{stokhof2020using, yang2024enhancing, hwang2021multidimensional}. We organize characters, images, and code concepts into nodes within the mind map to help students form a structured understanding. Inspired by construct-on-scaffold mind mapping \cite{zhao2022effect}, MindScratch uses three types of nodes to organize the programming project's mind map. These three types of nodes are distinguished by different colors: characters (green), logic (purple), and code (gray). We adopted a Scratch-style block palette (Figure \ref{fig:UI}.(a.4)) that allows students to select and add Scratch blocks. 

Another advantage of using mind maps is that we can provide more explanatory information when students use MindScratch. An additional LLM is instructed to annotate the generated content and serve as edges between nodes in the graph (Figure \ref{fig:UI}.(a.5)). To avoid information overload, MindScratch only shows brief relationship at a time, and students can click the relationship to check more. We introduced two features to address educators' challenges in using MindScratch to meet their clearly defined classroom objectives: setting up the initial mind map and defining objectives. Our system allows educators to construct mind maps as a starting point and input learning objectives to control the system's generation. When students are creating mind maps, we instruct an underlying LLM to ensure the generated content revolves around the learning objectives set by teachers. When students add nodes, our system detects whether the added nodes are relevant to the learning objectives and highlights those nodes if they are relevant (Figure \ref{fig:UI}.(a.6)). After completing the mind map, students can engage in hands-on creative programming based on the theme tasks outlined in the mind map. For example, students can use the code blocks from the mind map to implement specific features and import materials from the mind map. If they encounter issues during the programming process, they can still seek help from MindScratch for code suggestions. It is important to note that at no point does MindScratch generate a complete code solution directly; instead, it offers logic suggestions and key code blocks.

\subsubsection{Dialogic Question-Answering Box}

\begin{figure*}[htp]
    \centering
    \includegraphics[width=0.9\textwidth]{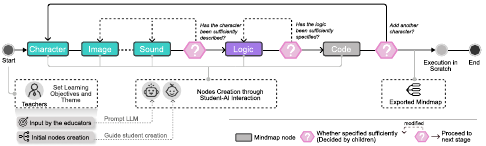}
    \vspace{-0.11in}
    \caption{\textbf{MindScratch user interaction process: Teachers set learning objectives and programming themes to prompt the LLM and guide student creation. Students collaborate with MindScratch to create characters, generate images and audio, and add them to the mind map. In the coding phase, students use LLM-provided logic and code blocks for implementation. Finally, students conduct hands-on Scratch programming and can ask MindScratch for help if needed.}}
    \label{fig:dialogue}
\end{figure*}

Based on prior research \cite{ruan2019quizbot,peng2022crebot}, we designed a dialogue interface to facilitate deeper student reflection and provide support during the project development process. The bottom of Figure \ref{fig:dialogue} illustrates how an educator sets programming themes and prompts learning objectives in MindScratch at the beginning of the class. The LLMs in MindScratch use the established learning objectives as memory, incorporating more content related to these objectives in their generation, such as highlighting mind map nodes or providing additional explanations. The top of Figure \ref{fig:dialogue} describes the interactive process between students and MindScratch. In the conversation process, the system captures students' needs and generates suggestions in a multiple-choice format. For example, the system guides students to think about the required characters by asking: \textit{``What characters will your project include?''} Students can then choose to manually add nodes or click the "character generation" button to select suggested characters for their mind maps. In the materials creation stage, if students click on a character node (marked in green), the system inquires, \textit{``Would you like to add an image or sound to this interesting character?''} This interaction enables students to transition to an AI-driven drawing board or a text-to-audio input box, facilitating the creation of multimedia materials. During the code implementation stage, MindScratch provides scaffolding to encourage students to elaborate their programming ideas. Students can receive assistance by clicking on the ``generate logic'' or ``generate code'' button within the chat box. For instance, if the child provides a brief description of a character without much detail, the agent will ask a follow-up question, \textit{``Very good! Would you like to add an action to the character?''} If the child is still confused, they can click the ``generate logic'' button in the chat window, and the system will generate various programming logic options for them to choose from (Figure \ref{fig:UI}.(b.1)). This feature particularly benefits beginners who may not fully grasp how algorithmic logic translates into actual Scratch code. For further clarification and assistance, the ``generate code'' button within the chat window allows students to generate Scratch code (Figure \ref{fig:UI}.(b.2)).

\subsubsection{AI-driven Drawing Board}

\begin{figure}[htp]
    \centering
\includegraphics[width=0.46\textwidth]{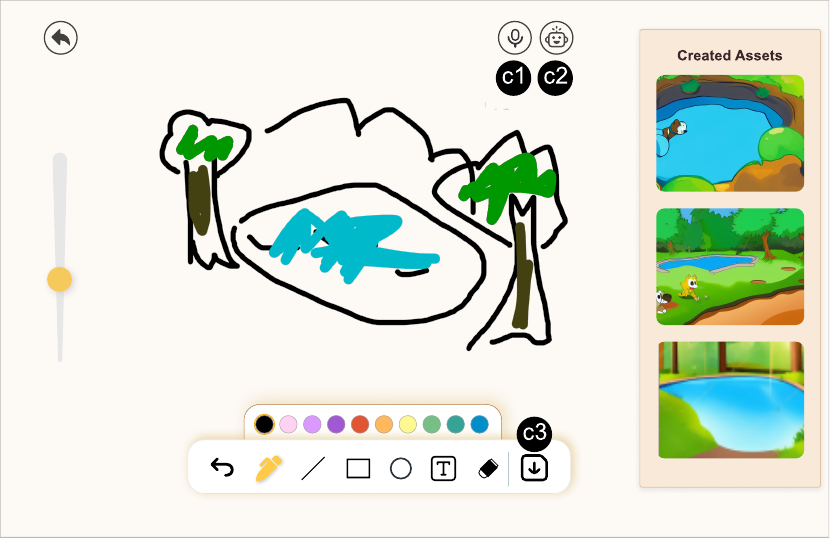}
    \vspace{-0.11in}
    \caption{\textbf{MindScratch's drawing board. It enables students to create, refine, and save image materials (c1, c2, c3).}}
    \vspace{-0.11in}
    \label{fig:UI2}
\end{figure}

To facilitate the easy creation of elements, such as characters and scenes in the mind map, MindScratch offers an AI-driven drawing board for sketching and refining images (Figure \ref{fig:UI2}). Students can utilize drawing tools to sketch, erase, select colors, and adjust line widths on the canvas. Given students' limited drawing skills, the system allows them to draw only basic sketches. By pressing the voice button (Figure \ref{fig:UI2}.(c.1)) to input speech and clicking the "image polish" button (Figure \ref{fig:UI2}.(c.2)), the system automatically refines their sketches into high-quality image assets. Students can regenerate images iteratively until they are satisfied with the output. Students can save their sketches (Figure \ref{fig:UI2}.(c.3)) or refined images, which will be displayed as thumbnails in the "created assets" column on the right. We also emphasize enabling students to create audio assets to enrich their projects. Once images are saved to the mind map, students can click the "sound generation" button and input text to generate interesting sound effects.

\subsection{MindScratch's Backend Model}

\subsubsection{Stage-based Dialogue Process.}

MindScratch uses dialogue across three phases—project planning, material creation, and code implementation—to support students in developing creative programming projects. The dialogue is powered by an LLM. To ensure the LLM provides guidance based on learning objectives, we use the teacher's input as a system prompt to instruct the LLM to follow these learning goals when generating logic and code suggestions. As long and complex prompts tend to decrease the task performance of LLMs \cite{wu2022ai, brown2020language}, we used dedicated prompts for each stage instead of combining instructions for all phases in a single prompt. Figure \ref{fig:dialogue} illustrates the stage-based dialog process for prompting LLMs. We instruct GPT-4 to support three primary tasks consistent with the classroom process: 1) Crafting questions that guide students in breaking down tasks, such as conceptualizing characters, actions, and events. 2) Formulating questions that help students create materials, including sketching and audio generation. 3) Offering help with programming logic or generating code, tailored to the students' input. To ensure the consistency of the content generated by LLM, we use the mind map information as a memory to prompt the LLM at every stage.

\subsubsection{Relationship Annotation and Block Highlighting.}

To enhance the interpretability of the content generated by LLMs, we employ few-shot prompting to instruct an additional GPT-3.5 in generating semantic relationships between nodes. We express it in triplets, such as <Theme, Relation, Character> and <Logic, Relation, Code>. However, annotating the relationships of all nodes might lead to the inclusion of less important relationships in the text, causing information overload for students. Therefore, we only label the edges connecting adjacent nodes generated by the LLM. To align the teacher's set learning objectives with the students' developed projects, we utilize the LLM to detect nodes related to the learning objectives within the mind map. The system then highlights these elements. For example, if the learning objective includes understanding conditional logic, the code blocks related to conditional logic will be highlighted. Drawing from prior work \cite{jiang2023graphologue}, we leverage saliency filters to manage the complexity of the mind map. We instruct GPT-4 to determine whether code blocks should be highlighted, designated as either \textit{high} (\$H) or \textit{low} (\$L). We only highlight code blocks that are highly relevant to the learning objectives, to reduce confusion among students.

\subsubsection{Multimodal Assets Generation.}

MindScratch integrated advanced image generation technologies based on the Stable Diffusion model \cite{StableDiffusion2023} and ControlNet \cite{zhang2023adding} to refine students' doodles. ControlNet incorporates an extra step by taking an input image and employing the Canny edge detector to identify its outlines. The resulting image, highlighting these detected edges, is saved as a control map. This control map is then used as supplementary conditioning alongside the text prompt when feeding into the ControlNet model. It is then fed to Stable Diffusion as an extra conditioning together with the text prompt. The refined images are generated based on these two conditionings. MindScratch employed a text-to-audio model \cite{ghosal2023tango} based on the diffusion model to generate high-quality audio materials. It is worth noting that we instruct GPT-4 to transform students' inputs into prompts that align with both the image and audio generation models. 

We conducted an image quality experiment to ensure that the images generated by MindScratch meet children's needs. The results show that the images created by MindScratch exhibit significantly better alignment with built-in materials compared to community-sourced assets. To quantify this, we used the Fréchet Inception Distance (FID) and Sliced Wasserstein Distance (SWD), which assess stylistic, textural, and structural similarities between image sets. Lower scores on both metrics indicate greater resemblance. As a baseline, the FID and SWD between built-in and community-sourced materials are 48.40 and 126.40, respectively. In contrast, the difference between MindScratch-generated assets and the built-in library is much smaller, with an FID of 8.12 and an SWD of 62.30. Additionally, we have taken measures to ensure that the generated content is safe and appropriate for students. To this end, we use a role-playing strategy with the LLM \cite{shanahan2023role}, instructing it to act as both an educator and a child's assistant. We also implement negative prompts \cite{Kapoor2023NegativePrompts} in the stable diffusion model, explicitly excluding keywords related to violence, horror, sex, crime, and discrimination, thereby safeguarding the content for a young audience.

\subsubsection{Step-by-step Code Assistant with LLMs.}

\begin{figure*}[htp]
    \centering
    \includegraphics[width=0.85\textwidth]{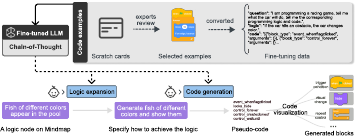}
    \vspace{-0.11in}
    \caption{\textbf{Overview of the logic-code assistance pipeline for MindScratch.}}
    \vspace{-0.11in}
    \label{fig:model}
\end{figure*}

Figure \ref{fig:model} shows the logic-code assistance pipeline for MindScratch. We proposed a scaffolded code support process, which provides step-by-step logic expansion and code generation support. To maintain consistency with the support pipeline and reduce the time required for generation, we fine-tuned a large language model. As shown in Figure \ref{fig:model}, we created 98 Scratch code samples from Scratch cards \footnote{\url{https://resources.scratch.mit.edu/www/cards/en/scratch-cards-all.pdf}}. To fine-tune the LLM, we converted the Scratch code samples into pseudo-code form. We adopted chain-of-thought reasoning to control the LLM's output and enhance the quality of the response. Since the generated code is in pseudo-code form, we further visualize the generated code into the mind map. Specific details are described in the following subsections.

\textbf{LLM fine-tuning.} We employ the GPT-3.5-turbo model as the basis for our fine-tuning process. Initially, we curate a dataset specifically tailored to our objectives, consisting of examples that closely mimic the types of interactions and outputs we aim for the model to replicate. This dataset includes a mix of dialogues, coding logic descriptions, and code solutions. The data collection and processing process is shown at the top of Figure \ref{fig:model}. In the Scratch community, Scratch cards emerge as essential introductory materials for programming. To obtain high-quality data for fine-tuning, we hired three educators, each with over four years of experience in teaching Scratch. Their task was to independently review all Scratch examples and select samples of educational significance. After an hour of review, educators selected 98 representative code examples. Subsequently, two researchers compiled these code examples into the fine-tuning dataset in the chain-of-thought style. BLEU and F1-score metrics are added to evaluate the performance of the LLM. Using 30 test samples based on Scratch cards, the fine-tuned LLM achieved a BLEU score of 32.1 and an F1 score of 80.9, compared to the vanilla LLM's BLEU score of 29.5 and F1 score of 73.2.

\textbf{Prompt Engineering.} Table \ref{tab:prompt} displays examples of prompt engineering for logic and code generation. A concrete example aids LLMs in eliciting the specific knowledge and abstractions required to complete a task \cite{yao2022react, wei2022chain}. To ensure that the fine-tuned LLM adheres to our code assistance workflow, we transformed the collected code examples into logic-code pairs and formulated the instruction in a chain-of-thought style. During the testing by researchers, it was discovered that although LLMs can generate correct pseudo-code, there remain issues with parsing, especially with nested relationships, which leads to decreased match performance and, consequently, an inability to visualize. We instruct the fine-tuned LLM to generate an abstract syntax tree (AST), which preserves the hierarchical structure of the code, thereby facilitating more precise alignment.

\textbf{Visualization of Generated Code.} To convert the pseudo-code generated by the LLM into visual Scratch code for easier understanding by students, we collected screenshots and class IDs of all Scratch code blocks from the Scratch website. We use the edit distance to determine how far the pseudo-code block generated by the LLM is from the correct code block's class ID. We parse all the pseudo-code blocks from the AST generated by the LLM, then calculate the edit distance with all class IDs, and assign the class ID with the smallest distance as the generated block ID. For example, if the LLM outputs the pseudo-code ``sensing touching\_object Cat'', it can be matched to the corresponding Scratch block ID ``sensing\_touchingobject''. This matching approach also helps to minimize errors caused by hallucinations produced by LLMs \cite{yao2023llm}. We then use the matched class ID to index the image of the code block, thereby visualizing the code block the students want, helping them correspond more easily with Scratch.

\subsection{Implementation}
MindScratch's mind maps are implemented using JSMind \footnote{\url{https://github.com/hizzgdev/jsmind}}, and the user interface is built with Vue.js \footnote{\url{https://github.com/vuejs/vue}}. The backend framework was developed using Flask \footnote{\url{https://github.com/pallets/flask}} in Python. We have utilized OpenAI's Whisper model to generate audio outputs for the conversational agent, and OpenAI's speech recognition API for transcribing speech \footnote{\url{https://openai.com/research/whisper}}. To minimize the interference of simultaneous voices in the classroom, we also designed a typing interface. To run the underlying LLMs, we used OpenAI's ChatCompletion API \footnote{\url{https://platform.openai.com/docs/guides/gpt/chat-completions-api}}. After extensive testing, we chose the state-of-the-art GPT-4-0613 model to generate responses and translate students' prompts to the image and audio generation model. In terms of safety, we incorporated a moderation layer \footnote{\url{https://platform.openai.com/docs/api-reference/moderations}} to ensure outputs were devoid of sexual content, hate speech, harassment, violence, or self-harm. The final version of prompts used in our system is displayed in our supplementary materials.

%% file: sections/exp.tex
\section{EXPERIMENT}
To investigate MindScratch's impact on students' creative programming process and outcome, we conducted a within-subject study, using the original Scratch platform as a baseline. Twenty-four fifth-grade elementary school students participated in our experiment. Each student used both systems, executing theme-based creative visual programming tasks in two separate sessions. Primarily, we aimed to address two research questions through our experiments:

\begin{itemize}

\item RQ1: How does MindScratch facilitate students' learning objectives in creative programming courses, reflecting the knowledge points and key tasks established by teachers?
\item RQ2: How does MindScratch enhance the students' project code quality and computational thinking skills in creative programming courses?
\item RQ3: How does MindScratch provide creative support for students in creative programming courses?
\item RQ4: What are the perspectives of educators regarding learner-focused AI assistants like MindScratch in terms of its integration into the curriculum, recommendations for improvement, and effective pedagogy?
\end{itemize}

\subsection{Participants}
This study took place in a primary school classroom in Shenzhen, China, during the winter of 2023, involving 24 fifth-grade students. The participants were 24 fifth-grade students. All participants had already completed a semester-long Scratch course. All participants had a basic understanding of how to use Scratch's fundamental code blocks, as well as basic software operations such as uploading and editing images, and uploading and editing audio. However, they were not familiar with advanced CT skills, such as nested condition loops and the broadcast mechanism. To prevent the potential influence of different teachers on the outcome of the study, all students were taught by the same teacher using the same technical equipment regardless of which group they belonged to. The classes were taught by a teacher with 6 years of experience in teaching Scratch courses. The participants' ages ranged from 10 to 11 years. Each student engaged with two systems, Scratch and MindScratch, completing two theme-based creative programming tasks across separate sessions. All participants were native Mandarin speakers. As compensation for participating in the experiment, each student received a gift worth \$15.

\renewcommand{\arraystretch}{1.1}
\begin{table*}
\footnotesize
\centering
\caption{\textbf{Examples of prompt engineering for code generation.}}
\vspace{-0.11in}
	\label{tab:prompt}
	\begin{tabularx}{\linewidth}{>{\bfseries} l|X} 
		\hline
		\rowcolor{darkgray} \bf{PROMPT TYPE}  & \bf{INSTANTIATION} \\
		\hline
            \hline
            \multicolumn{2}{l}{\textbf{Example of logic generation}} \\
            \midrule
		LLM instruction & \textit{\makecell[Xt]{Your task is to generate a programming logic node based on the provided mind map and corresponding question. The generated programming logic must not repeat the content of the mind map; the description must be concise and suitable for a child's level. The response should be in JSON format. Below are some examples.}} \\
		  \midrule
    	The input example in LLM & \textit{\makecell[Xt]{\text{Mind map:} \{
    \text{"racing game":} \{
        \text{"car":} \{
            \text{"function":} ["Keep the car moving", "Adjust the car's direction"]
        \},
        \text{"track":} \{
            \text{"function":} ["Touching the track causes the track to change color"]
        \}
    \}
\}}} \\
		\midrule
		User's input & \textit{Car.} \\
		\midrule
        The output example in LLM & \textit{\{"logic": Adjust the car's direction.\}}  \\
            \midrule
            \addlinespace[0.1cm] 
            \multicolumn{2}{l}{\textbf{Example of code generation}} \\
            \midrule
		LLM instruction & \textit{\makecell[Xt]{Please provide well-structured Scratch code based on the programming logic entered by the user. Ensure that the generated code blocks can be correctly parsed by Scratch software. The response should be in JSON format. Below are some examples.}} \\
		  \midrule
    	The input example in LLM & \textit{Here is an example: \{ "code": [ \{"block\_type": "event\_whenflagclicked", "arguments": \{\}\}, \{"block\_type": "control\_forever", "arguments": \{\}\}, \{"block\_type": "control\_if", "arguments": \{ "condition": \{ "block\_type": "sensing\_touchingobject", "arguments": \{ "object": "obstacle" \}\}, "block\_type": "looks\_seteffectto", "arguments": \{ "effect": "color", "value": 100 \}\}, "block\_type": "operator\_subtract", "arguments": \{ "NUM1": \{ "block\_type": "sensing\_of", "arguments": \{ "property": "score", "object": "Stage" \}\}, "NUM2": 1 \}\}\}\} ] \}}\\
		\midrule
		User's input & \textit{\makecell[Xt]{Keep the car moving.}} \\
		\midrule

        The output example in LLM & \textit{\makecell[Xt]{[\{"block\_type": "event\_whenflagclicked", "arguments": \{\}\}, \{"block\_type": "control\_forever", "arguments": \{\}\}, \{"block\_type": "motion\_movesteps", "arguments": \{"steps": 10\}\}]}}  \\
        \bottomrule
	\end{tabularx}

\end{table*}
\subsection{Procedure}

As a within-subject study, each participant was required to undertake two theme-based creative programming sessions. To avoid potential time-related effects and carryover effects, we introduced a one-week interval between sessions. Moreover, we employed a counterbalanced design \cite{davier2004kernel}, alternating between combinations of tools (MindScratch and Scratch) and themes (A or B) to ensure the fairness of the outcomes. Inspired by previous studies and the insights of educators \cite{mueller2017assessing,zhang2022storydrawer}, we and the teacher designed two creative programming tasks: A) ``Cat and Mouse'' and B) ``Kitten Fishing''. Each session lasted about 80 minutes, with 20 minutes dedicated specifically to familiarizing with the MindScratch system and free exploration. Participants then had 10 minutes to ask questions, which were answered by researchers. The teacher spent 20 minutes explaining the key points of the programming task. The last 30 minutes were focused on the theme-based creative programming task. The specific details of students' theme-based creative programming tasks can be found in Appendix A. MindScratch and Scratch were both configured with a Chinese interface according to the participants' preferences. During the experiment, researchers were only allowed to assist the students in the event of unexpected technical difficulties. After each session, we collected screen recordings through capture software. Additionally, we collected the project files created by the students in sb3 format. Subsequently, the participants were asked to complete a creativity support index questionnaire \cite{cherry2014quantifying} to evaluate how much effective creative support their process received. After completing these two parts, we conducted brief semi-structured interviews with the participants to gain deeper insights into their user experience with MindScratch for class-based creative programming tasks.

We also explored the long-term effects of MindScratch on the development of computational thinking in students. Three students from the class participated in an experiment during the winter vacation. We invited the teacher to use our tools to conduct a four-week one-on-three course, with one lesson per week. Each lesson involved programming one project using MindScratch. Before and after the course started, we conducted pre- and post-tests on them using a CT skills survey \cite{korkmaz2019adapting}. After the course ended, we asked them for their feedback and opinions on how the tool supported their programming in the classroom. 

To further understand how educators utilize MindScratch in creative programming courses, we conducted semi-structured interviews with six educators in graphical programming. The interviews began with exploring the educators' backgrounds and their current challenges and strategies, particularly regarding students' creative exploration and programming implementation using LLM-based tools. Subsequently, we introduced the creative programming support assistant, discussed its capabilities, and the insights gathered from its long-term deployment, as summarized in Sections 6.1-6.3. The discussion then shifted to educators' perceptions of our tool: what they liked and disliked, their pedagogical and ethical considerations for using it, their interest and requirements for integrating it into their curriculum, and how they view its relationship with tools like ChatGPT. Each interview was conducted on Tencent Meeting and lasted about 1 hour.

\subsection{Measurements}
In this section, we describe the evaluation process, including the six categories of data collected for assessment. Table \ref{tab:Collected Data and Metrics Summary} displays the metrics used for quantitative assessment. Interviews based on artifacts were used for qualitative insights. Any data requiring subjective evaluation were independently assessed by multiple evaluators, with their agreement documented.

\begin{table*}[htp]
\centering
\footnotesize
\caption{\textbf{Summary of the collected data and the measures used in the experiments.}}
\vspace{-0.11in}
\label{tab:Collected Data and Metrics Summary}
\begin{tabularx}{0.95\textwidth}{p{0.23\textwidth}p{0.23\textwidth}X}
\toprule
Data & Evaluation Metrics & Description \\
\midrule
Code Quality Scores& Dr.Scratch Rubric& Assess code quality by quantifying seven
computational thinking dimensions \cite{moreno2015dr}.  \\
\addlinespace
Expert Ratings& Expert Ratings on Projects & Evaluation by Scratch educators to determine the logic consistency, quality, creativity, and originality of students' projects \cite{amabile1982social}. \\
\addlinespace
Mind Map Node Count& Mind Map Richness & Count of graph nodes in projects to quantify project richness \cite{kovalkov2021automatic}.\\
\addlinespace
Creativity Support Index Questionnaires& Creativity Support Index & Measure the usability and effectiveness of a system in enhancing creative programming tasks \cite{cherry2014quantifying}. \\
\addlinespace
CT Skills Pre and Post-Tests& CT Skills Survey & Measure the long-term potential impact of the system on the cultivation of computational thinking \cite{korkmaz2019adapting}. \\
\addlinespace
Artifact-based Interview& Semi-Structured Interview & Gather insights on the creation of students' mind maps, classroom progress, and feedback. \cite{portelance2015code}. \\
\bottomrule
\end{tabularx}

\end{table*}

\subsubsection{Dr.Scratch Rubric Scores.} 
We adopted the Dr.Scratch scoring criteria \cite{moreno2015dr} to assess the code quality of students' class-based programming projects. As a widely applied metric for evaluating Scratch code, it assesses code quality by quantifying seven computational thinking dimensions: abstraction, parallelism, logic, synchronization, flow control, interactivity, and data representation. The score for each dimension ranges from 0 to 3, representing three levels: basic, developing, and master. Essentially, higher code quality scores indicate that students have more extensively utilized and mastered CT skills during the project development process \cite{dietz2023visual}. 

\subsubsection{Expert Ratings.} For this evaluation, three Scratch education experts were invited to assess the students' projects independently. The experts used a 5-point Likert scale based on four criteria adapting from \cite{amabile1982social} for evaluating the projects: (1) Consistency: indicating the extent to which a student's projects reflect the classroom learning objectives; (2) Quality: reflecting the intrinsic properties of the project, including completeness, clarity, detail, and harmony; (3) Originality: indicating the extent to which the project reflects the student's creation rather than being derived from existing materials; (4) Creativity: indicating innovative content expression and rich imagination in the project's materials. The evaluation results showed a high degree of consensus among the experts, with an overall intraclass correlation coefficient (ICC) of 0.78 (p < 0.05).

\subsubsection{Creativity Support Index.} 
The creativity support index (CSI) \cite{cherry2014quantifying} measures users' perceptions of tool usability and support for creative tasks. This survey consists of 12 questions distributed across six themes. We employ a 5-point Likert scale to collect feedback and ensure each student is in an undisturbed environment, allowing us to gather firsthand feedback from the students and gain deeper insights into their experiences. 

\subsubsection{Mind map Node Count.} In the classroom, teachers' ideas often influence students' project planning. Inspired by \cite{kovalkov2021automatic}, we adopted graph node count to reflect the richness of the mind maps used for programming. In our evaluation, the number of graphical nodes is counted as the sum of character nodes (project planning and multimodal materials) and programming nodes (logic and code blocks).

\subsubsection{CT Skills Survey.} To investigate the long-term impact of MindScratch on students' CT skills, we employ a 5-point Likert scale revised CT skills survey proposed by \cite{korkmaz2019adapting}, which includes creativity (4 items), cooperativity (4 items), critical thinking (4 items), algorithmic thinking (4 items), and problem-solving (4 items). Before the four-week course began, we conducted a pre-test on the students using the CT skills survey, and a post-test was administered after the course concluded.

\subsubsection{Artifact-based Interview.}
We conducted interviews using two artifacts: the mind maps and the projects created. This allows us to gain a deeper understanding of whether MindScratch effectively supports learning objectives and creative programming. We structured our interview around three themes to understand the students' projects created around the classroom objectives provided by the teacher (questions 1-2), their classroom process (questions 3-4), and their views on MindScratch (questions 5-7):

\begin{enumerate}
    \item Tell me about your projects.
    \item Did you follow the classroom objectives? Was your idea realized?
    \item We noticed that during [specific time/event], you exhibited [specific behavior]. What problems did you encounter at that time?
    \item How did you address these challenges?
    \item Would you be willing to use MindScratch in class?
    \item Which features of MindScratch do you find useful?
    \item Do you think using MindScratch would make programming easier for you?
\end{enumerate}

\subsection{Data Analysis}

We implemented a counterbalance design in our experiment to mitigate the effect of order and theme on the programming outcomes. The paired t-test was used to investigate the differences between Scratch and MindScratch. A Bonferroni correction is applied to mitigate the risk of Type I errors associated with multiple comparisons. All the data and analysis codes are included, for details, please see our supplementary materials.

%% file: sections/results.tex
\section{results}

\subsection{RQ1: Impact on Achieving Learning Objectives}

\begin{figure}[thp]
    \centering
    \includegraphics[width=0.48\textwidth]{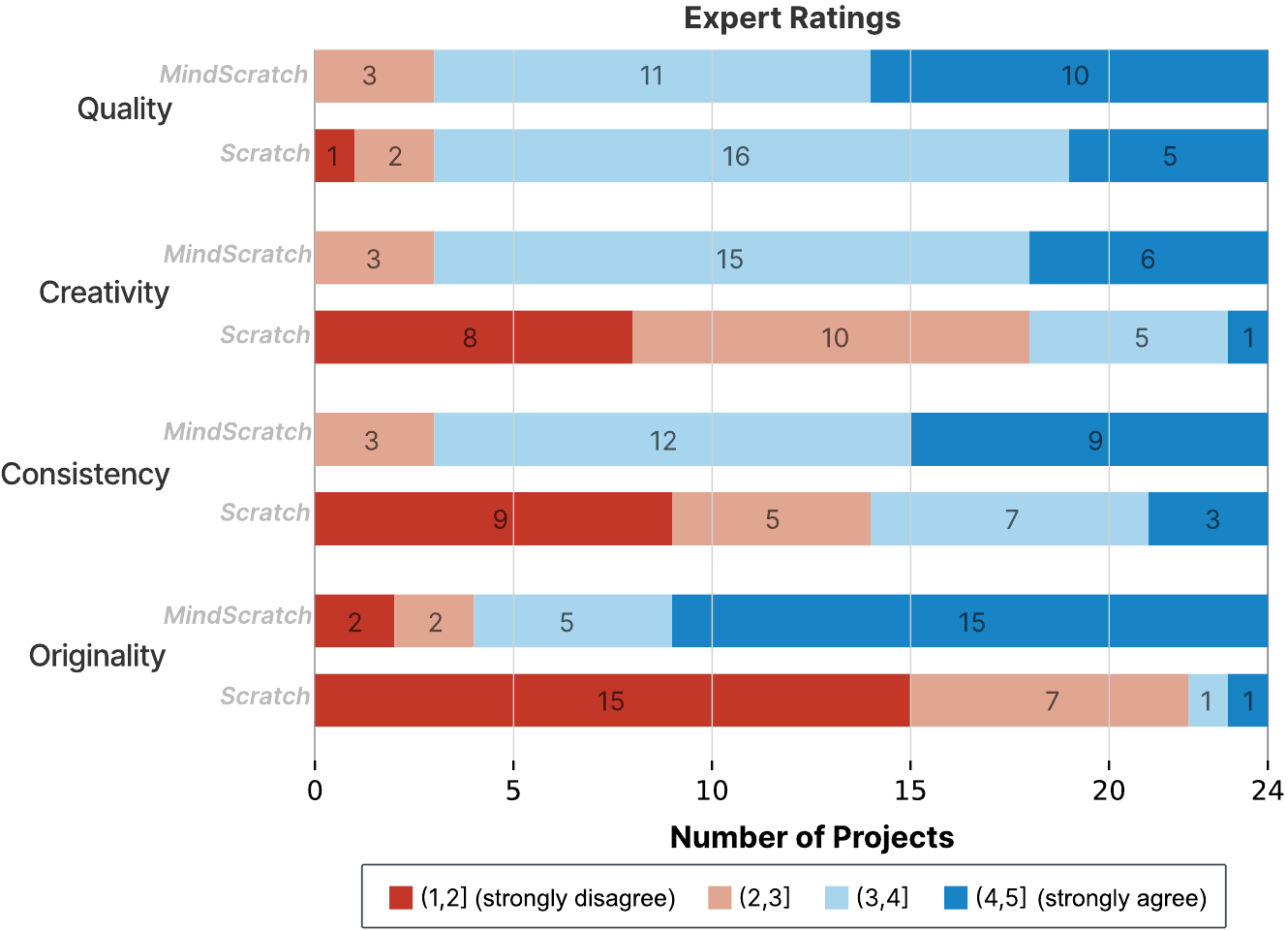}
    \vspace{-0.11in}
    \caption{\textbf{The score distribution between MindScratch and Scratch. Note that higher values indicate positive feedback, and vice versa.}}
    \vspace{-0.11in}
    \label{fig:expert_rating}
\end{figure}

Figure \ref{fig:expert_rating} shows the distribution of scores by experts. In the same class time, students using MindScratch outperformed those using Scratch in terms of consistency, originality, and creativity. All 24 students who used MindScratch completed the teacher's pre-defined tasks and learning objectives. For instance, the objectives set by the teacher included making a cat follow the mouse cursor, having multiple mice appear, and each mouse moving continuously, among others. However, only 13 students using Scratch achieved the learning objectives. Although many projects created by students using Scratch contained interesting interactions, it is clear that these projects did not reflect the classroom learning objectives. Without timely intervention from the teacher, these projects would lack completeness. In particular, Table \ref{tab:expert_rate} shows that MindScratch plays an important role in promoting the consistency between programming projects and learning objectives \((t(23) = 5.32, p < 0.01^{**})\). We found that experts appreciated the projects created using MindScratch. Significant differences were noted in originality $(t(23) = 6.86, p < 0.01^{**})$ and creativity $(t(23) = 7.60, p < 0.01^{**})$ between MindScratch and Scratch. For instance, P15 used MindScratch to craft a narrative about a fortunate mouse and a less fortunate cat. In the story, when the owner is away, the mouse ingeniously disguises itself as a lion to frighten the cat away. This story not only reflects the classroom learning objectives but also includes rich and exquisite assets, showcasing the student's creativity effectively. Besides, in terms of quality, we observed matched means (4.10 for MindScratch and 4.04 for Scratch) and standard deviations (0.59 for MindScratch and 0.69 for Scratch), which suggested that projects created with MindScratch in the classroom have the same level of completeness as those made with Scratch.

\begin{table*}[thp]
    \centering
    \footnotesize
    \renewcommand{\arraystretch}{1.2}  
    \caption{\textbf{Comparative Evaluation of MindScratch vs. Scratch Across Expert Ratings.}}
    \vspace{-0.11in}
    \begin{tabularx}{\textwidth}{l *{2}{>{\centering\arraybackslash}X} c *{2}{>{\centering\arraybackslash}X} c *{2}{>{\centering\arraybackslash}X}}
        \toprule
        Metric & \multicolumn{2}{c}{MindScratch} & & \multicolumn{2}{c}{Scratch} & & \multicolumn{2}{c}{Paired-t test} \\
        \cline{2-3} \cline{5-6} \cline{8-9}
        & Mean & SD & & Mean & SD & & t & p \\
        \hline
        Originality & 4.38 & 0.97 & & 2.48 & 0.88 && 6.86 & $0.000^{**}$ \\
        Consistency & 4.25 & 0.67 & & 3.0 & 1.32 && 5.32 & $0.000^{**}$\\
        Creativity  & 4.13 & 0.61 & & 2.91 & 0.93 && 7.60 & $0.000^{**}$ \\
        Quality     & 4.10 & 0.59 & & 4.04 & 0.69 && 1.30 & 0.83 \\
        \bottomrule
    \end{tabularx}
    \begin{tablenotes}
    \footnotesize
    \item[] \textbf{Notes:}
    \item 1. A Bonferroni correction is applied to mitigate the risk of Type I errors associated with multiple comparisons.
    \item 2. ** denotes $p < 0.01$ and * denotes $p < 0.05$.
    \end{tablenotes}
    \label{tab:expert_rate}
\end{table*}

\subsection{RQ2: Effectiveness on Code Quality and CT Skills}
\begin{table*}[thp]
    \centering
    \footnotesize
    \renewcommand{\arraystretch}{1.1}  
    \caption{\textbf{Comparative Evaluation of MindScratch vs. Scratch Across Dr. Scratch Rubrics.}}
    \vspace{-0.11in}
    \begin{tabularx}{\textwidth}{l *{2}{>{\centering\arraybackslash}X} c *{2}{>{\centering\arraybackslash}X} c *{2}{>{\centering\arraybackslash}X}}
        \hline
        Metric & \multicolumn{2}{c}{MindScratch} & & \multicolumn{2}{c}{Scratch} & & \multicolumn{2}{c}{Paired-t test} \\
        \cline{2-3} \cline{5-6} \cline{8-9}
        & Mean & SD & & Mean & SD & & t & p \\
        \hline
        Abstraction & 2.17 & 0.56 & & 1.71 & 0.55 & & 3.11 & $0.039^{*}$ \\
        Parallelism & 2.19 & 0.51 & & 1.58 & 0.50 & & 3.42 & $0.011^{*}$ \\
        Logical & 2.04 & 0.81 & & 1.13 & 0.34 & & 5.79 & $0.000^{**}$ \\
        Synchronization & 2.08 & 0.95 & & 1.46 & 0.66 & & 3.50 & $0.016^{*}$ \\
        Flow Control & 1.92 & 0.50 & & 1.63 & 0.49 & & 2.07 & 0.399 \\
        Interactivity & 2.13 & 0.61 & & 1.54 & 0.51 & & 3.98 & $0.005^{**}$ \\
        Data & 2.04 & 0.20 & & 0.91 & 1.41 & & 3.02 & $0.049^{*}$ \\
        \hline 
        Total Score & 14.17 & 4.14 & & 9.96 & 4.46 & & 6.44 & $0.000^{**}$ \\
        \hline
    \end{tabularx}
    \begin{tablenotes}
    \footnotesize
    \item[] \textbf{Notes:}
      \item 1. A Bonferroni correction is applied to mitigate the risk of Type I errors associated with multiple comparisons.    
    \item 2. ** denotes $p < 0.01$ and * denotes $p < 0.05$.
    \end{tablenotes}
    \label{tab:dr_scratch_rubrics}
\end{table*}

Table \ref{tab:dr_scratch_rubrics} shows the results of the Dr.Scratch rubric, which evaluates the quality of code by assessing the level of CT in students' projects. For the total score, the mean for MindScratch was 14.17 (\(SD=4.14)\)), while that for Scratch was 9.96 (\(SD=4.46\)). Within the framework of Dr.Scratch, this progression marks the elevation of students' projects from a ``basic'' to a ``master'' level. Based on the paired-samples t-test, we observed an improvement in six dimensions: abstraction $(t(23) = 3.11, p < 0.01^{**})$, parallelism $(t(23) = 3.42, p = 0.011^{*})$, logical $(t(23) = 5.79, p < 0.01^{**})$, synchronization $(t(23) = 3.50, p = 0.016^{*})$, interactivity $(t(23) = 3.98, p < 0.01^{**})$, and data $(t(23) = 3.02, p = 0.049^{*})$, except flow control $(t(23) = 2.07, p = 0.399)$. These results suggest that compared to the baseline tool, MindScratch led to a better programming outcome in terms of code quality and logic organization. In the interview, ten students indicated that the interactive mind map and the scaffolded logic-code generation contributed to their programming output. \textit{``During the use of MindScratch, the code suggestions it gave me were helpful when I didn't know which code blocks to use to implement the logic''} (P14). \textit{``With MindScratch, I could figure out the logic for implementing code without needing help from a teacher''} (P2). We do not observe a significant effect on the difference in flow control. According to students’ feedback, we suppose that this can be attributed to the understanding and application of complex conditional logic, which requires a higher level of computational thinking \cite{selby2013computational}.

\begin{figure*}[thp]
    \centering
    \includegraphics[width=0.95\textwidth]{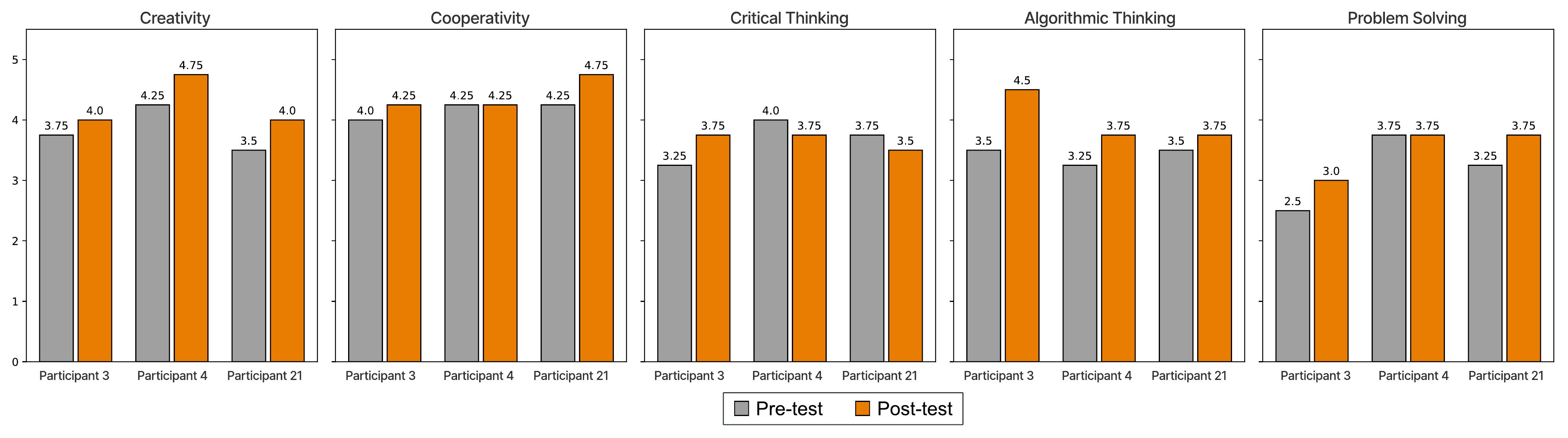}
    \vspace{-0.11in}
    \caption{\textbf{The bar plots illustrate the pre-test and post-test results of three students across five sub-dimensions of the CT skills survey.}}
    \label{fig:CtTest}
    \vspace{-0.11in}
\end{figure*}

Figure \ref{fig:CtTest} illustrates the pre-test and post-test results using the CT Skills Survey. Each student's CT skills improved, with P3's pre-test mean at 3.4 and post-test mean at 3.9, P4's pre-test mean at 3.9 and post-test mean at 4.05, and P21's pre-test mean at 3.65 and post-test mean at 3.95. Overall, there was an improvement in students' creativity, cooperativity, algorithmic thinking, and problem-solving. All three students mentioned they could more easily understand the relationships between code blocks (algorithmic thinking) and could design a complex programming plan (problem-solving). P3 stated, \textit{``After using MindScratch, I learned \textbf{how to use the code from previous projects} to solve current problems.''} P21 mentioned, \textit{``I now like to create a mind map to help me \textbf{conceptualize programming}; with a mind map, I can complete a project faster.''} We can observe that the creativity of the three students and the collaboration skills of the two students have been improved. P4 commented, \textit{``When creating programming projects, I want to use a variety of materials (images, audio). But each time, the materials provided by the teacher are limited. MindScratch allows me to access the materials I want more quickly.''} MindScratch does not show significance in the development of critical thinking. In future work, we plan to incorporate strategies from \cite{yuan2023critrainer} for critical thinking training to nurture this essential skill.

\subsection{RQ3: Effectiveness on Creativity Support}

\begin{figure}[thp]
    \centering
    \includegraphics[width=0.48\textwidth]{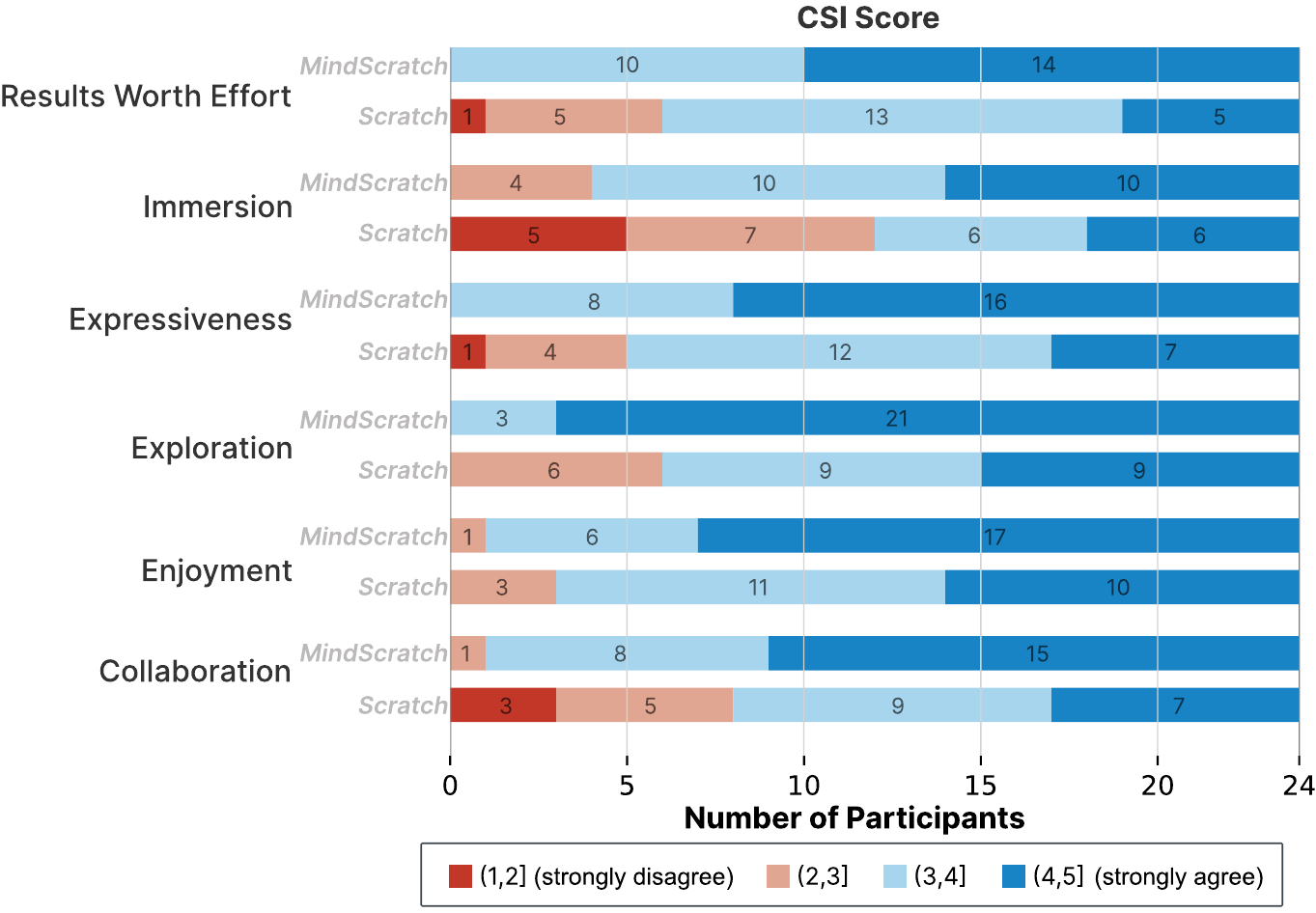}
    \vspace{-0.11in}
    \caption{\textbf{The score distribution between MindScratch and Scratch. Note that higher values indicate positive feedback, and vice versa.}}
    \vspace{-0.11in}
    \label{fig:csi_rating}
\end{figure}

\begin{table*}[thp]
    \centering
    \footnotesize
    \renewcommand{\arraystretch}{1.2} 
    \caption{\textbf{Comparative Evaluation of MindScratch vs. Scratch Across Creativity Support Index.}}
    \vspace{-0.11in}
    \begin{tabularx}{\textwidth}{l *{2}{>{\centering\arraybackslash}X} c *{2}{>{\centering\arraybackslash}X} c *{2}{>{\centering\arraybackslash}X}}
        \toprule
        Metric & \multicolumn{2}{c}{MindScratch} & & \multicolumn{2}{c}{Scratch} & & \multicolumn{2}{c}{Paired-t test} \\
        \cline{2-3} \cline{5-6} \cline{8-9}
        & Mean & SD & & Mean & SD & & t & p \\
        \hline
        Collaboration & 3.98 & 0.99 & & 2.83 & 0.89 & & 3.88 & \(0.005^{**} \) \\
        Enjoyment & 3.73 & 0.69 & & 3.35 & 0.80 & & 1.72 & 0.59 \\
        Exploration & 3.90 & 0.97 & & 2.94 & 0.97 & & 3.04 & \(0.035^{*} \) \\
        Expressiveness & 3.94 & 0.86 & & 3.02 & 1.02 & & 3.52 & \(0.011^{*} \) \\
        Immersion & 3.67 & 0.93 & & 2.83 & 0.88 & & 2.92 & \(0.005^{**}\) \\
        Results Worth Effort & 4.19 & 0.83 & & 3.23 & 0.78 & & 4.11 & \(0.003^{**}\)\\
        \bottomrule
    \end{tabularx}
    \begin{tablenotes}
    \footnotesize
    \item[] \textbf{Notes:}
     \item 1. A Bonferroni correction is applied to mitigate the risk of Type I errors associated with multiple comparisons.  
    \item 2. ** denotes $p < 0.01$ and * denotes $p < 0.05$.
    \end{tablenotes}
    \label{tab:csi}
    \vspace{-0.11in}
\end{table*}

Another objective of our study focuses on how MindScratch supports students in enhancing their creativity. First, We employed the metric of mind map node count to evaluate the students' efforts in classroom-based programming. Based on the learning objectives, the teacher created two complete mind maps as the baseline based on the two themes. The average count in MindScratch was 52.45 (SD = 6.93), while the baseline average was 35. This fully demonstrates that students can express their creativity with MindScratch based on the learning objective. 

Second, We also report on students' subjective perceptions of the creative support they received during the programming process. From Figure \ref{fig:csi_rating}, we can see that students have a positive view of both systems and MindScratch received higher scores across all six dimensions. A paired-sample t-test shown in Table \ref{tab:csi} suggests that MindScratch showed significant improvement in areas such as collaboration $(t(23) = 3.88, p < 0.01^{**})$, exploration $(t(23) = 3.04, p = 0.035^{**})$, expressiveness $(t(23) = 3.52, p = 0.011^{*})$, immersion $(t(23) = 2.92, p < 0.01^{**})$ and results worth effort $(t(23) = 4.11, p < 0.01^{**})$ compared to Scratch. This indicates that MindScratch provided better creative support to students while engaging in classroom project programming, enhancing their learning experience. Statistically, 12 out of 24 students mentioned the contribution of character generation, 21 highlighted the impact of image polish, and 8 acknowledged the audio generation. On average, the students used character generation 5.32 times (SD = 1.56), image polish 6.27 times (SD = 2.74), and audio generation 2.34 times (SD = 0.38). P23 stated, \textit{``It made classroom learning more interesting for me. I could create funny images and sounds on my own, which were different from what the teacher provided. When I was unsure about how to draw the characters, it also offered help.''}  In conclusion, by interactively creating mind maps (collaboration), students' efforts in creative tasks become more meaningful (results worth the effort). The visualization based on mind maps aids in exploring the details of projects (exploration), while multimodal asset creation enables students to effectively express their inspiration (expressiveness). Scaffolded logic-code support offers an efficient pipeline that allows for conceptualization and coding to be completed in one go. This enables students to focus more on creative programming (immersion).

\subsection{RQ4: Educator’s Perspectives}
To gain further insights into how educators would use MindScratch in their visual programming classes, we conducted semi-structured interviews with six teachers (T1 - T6). Their professional opinions provided more evidence on the appropriateness and effectiveness of using MindScratch in the classroom.

\subsubsection{General Impressions.}

Educators generally have a positive impression of MindScratch. T1 emphasized the pedagogical approach of MindScratch, describing it as offering a ``mind map creation method with AI collaboration,'' which is particularly helpful for students with unclear thoughts. Similarly, T3 mentioned that MindScratch is a more controllable way for elementary students to use large models,'' different from tools like ChatGPT that directly provide answers, instead, it guides students step-by-step to find answers. Furthermore, T5 envisioned that MindScratch could greatly assist teachers in dynamic creative classroom settings, helping teachers more easily handle spontaneous student issues, provide tailored feedback, and support teachers in quickly identifying students' knowledge gaps.

\subsubsection{Perceptions on Mind-map Usage.}
Most educators appreciate the design of the interactive mind map, especially how it provides explanations after students select nodes and highlight code blocks related to the set educational objectives. T6 liked that it facilitates creative exploration but helps students avoid excessive divergence. T2 and T3 mentioned how it reduces cognitive load by visualizing the overall logic of projects. T4 stated that ``providing logic-to-code guidance'' aids in students' metacognitive skills. However, T1 expressed concerns about the lack of support for advanced algorithms, as the generated code blocks struggle to reflect the characteristics of data structures.

\subsubsection{Perceptions on External Materials Generation.}
Most educators believe that using GAI to generate programming materials can enhance students' learning interests and stimulate their creativity. Moreover, T2 stated that having students create their materials reduces the teaching burden on teachers to support differentiated student needs. T4 believes that before using MindScratch, creative projects merely reflected the teacher's creativity, rather than projects meaningful to the students personally. T3 and T5 discussed how to enhance the relationship between material creation and classroom themes by combining teacher-set mind maps. However, T1 expressed slight concerns, particularly that students focus on drawing or debugging prompts to generate content, neglecting subsequent programming training.

\subsubsection{Concerns about Incorrect Responses and Misuses.}
Despite positive impressions, educators expressed various concerns. After viewing recordings of our system in use, T1 pointed out the risks of incorrect answers, especially for students with weaker foundations. T3 was concerned about students ``believing anything the large language model says.'' As a solution, T1 and T3 suggested including systematic courses and quizzes before students use MindScratch. T5 asked to improve the model's accuracy by having students identify incorrect answers and suggested that the system should highlight potential uncertainties, prompting students to reflect, such as ``Are you sure this code block will achieve the correct logic?'' Additionally, T1, T2, and T3 expressed concerns about students misusing MindScratch and suggested limiting the number of daily generations as a potential solution.

\subsubsection{Student Monitoring Dashboard.}

Another popular topic was the need for a teacher dashboard to monitor student interactions with the system and track their progress. T4 emphasized that by monitoring the questions students ask and the types of answers generated, educators can assess students' grasp of computational thinking, which can be obtained by the system generating reports on common issues. These insights could prompt interventions in the next class to provide better examples. However, accessing students' private data may be restricted. This raises a question about the balance between personalization and privacy. Instead of directly reporting student problems, the system should deal with potential privacy leaks first. T1 mentioned that students should not feel monitored while using the system.

%% file: sections/disccusion.tex

\section{Discussion}

In this section, we reflect on our findings from the development and evaluation of MindScratch. We also discuss future opportunities for research exploring visual programming support tools.

\subsection{Design Considerations}

\subsubsection{Enhancing generated Answer Reliability and Providing Reliable Sources}
Based on our interviews, educators are still concerned that students may believe everything the LLM says, even if the answers may be incorrect. Although MindScratch prompts the LLM to annotate the relationships between nodes to improve the interpretability of the generated content, potential errors are still unavoidable. Therefore, MindScratch should incorporate advanced Retrieval-Augmented Generation \cite{liu2024hita} to enhance the reliability of its responses, while also listing the sources referenced when providing answers. A high-quality programming knowledge base can greatly improve the quality and reliability of the LLM's generated responses.

\subsubsection{Balancing Support and Self-individual Effort.}
Drawing from the theory of scaffolding \cite{shabani2010vygotsky}, MindScratch adopts a question-answering format to encourage users to exert effort before receiving support and feedback. For example, after children create digital materials, the system poses questions to guide them in thinking about programming logic. During the programming phase, MindScratch provides a Scratch-style block palette to encourage children to think about code implementation on their own. Our user study indicates that children actively expand on the code suggestions provided by the system during the programming, reflecting their interests. However, the system does not take into account the cognitive and capability development of children, and we suggest that future visual programming assistance systems should provide support and feedback according to the user's programming level. Moreover, MindScratch could incorporate gamification features (e.g., leaderboards, and online project sharing) to encourage users to enrich their projects for better performance.

\subsubsection{Providing Teachers with Student Monitoring Dashboard.}
MindScratch visualizes students' thought processes during project creation through mind mapping, helping teachers monitor student progress and provide targeted feedback. However, educators believe that collecting additional process data (such as question data) would give a more comprehensive understanding of students' knowledge levels. MindScratch should offer a teacher dashboard to help educators track student progress and common questions. In future versions, MindScratch could further enhance collaboration between teachers and AI. For example, the system could allow teachers to set LLM prompts to provide assistance with varying levels of granularity based on classroom progress. Additionally, teachers could adjust the LLM’s knowledge level through a slider, giving them greater control over the system to tailor support according to students’ programming abilities.

\subsection{Limitations and Future Work}
Our research has several limitations. First, our user study mainly focused on a fixed age group of 10- to 11-year-olds, and the children in our experiment already had a preliminary understanding of Scratch and basic programming concepts. While this experimental setup provided us with a controlled environment, eliminating potential cognitive differences among children of different ages, it may also limit the system’s adaptability to children of other age groups, preventing us from identifying possible limitations. We encourage future studies to include children of various ages and interests, such as those aged 6 to 9.

Secondly, we compared the learning outcomes of MindScratch with traditional Scratch teaching. MindScratch leverages Generative AI technology to provide support at different stages of project development, thus improving learning outcomes. However, for children, the development of computational thinking is a long-term process. Currently, MindScratch does not offer assessments or feedback on completed programming projects, which limits its use as a self-learning tool. Future work could explore how to enhance MindScratch by integrating multi-perspective evaluation tools that assess aesthetics, code quality, and computational thinking skills \cite{chai2023towards}. Additionally, MindScratch currently supports only Scratch, and it may have limitations when applied to more complex text-based programming languages like Python. Future studies could explore extending MindScratch to multiple programming languages to increase the system’s applicability.

Finally, the safe use of AI by children has become a topic of public discussion. As we enter an era dominated by Generative AI, ethical considerations related to privacy and safety are increasingly important. For safety, we incorporated a moderation layer from OpenAI and prompt engineering to ensure that outputs contain no sexual content, hate speech, harassment, violence, or self-harm. However, this input-based filtering method may be influenced by biases in the training data. Future research could consider deploying large language models in specialized domains to limit the generation of uncontrollable responses, similar to our focus on programming support in this study. Such limitations could help reduce risks while still leveraging the capabilities of large language models.

%% file: sections/conclusion.tex
\section{Conclusion}

This paper introduces MindScratch, a visual programming support tool that assists students in completing creative exploration and programming implementation in the classroom. We are the first to propose a GAI-driven interactive mind map system to help students achieve learning goals and offer timely creative and programming support. The novelty of our research centers on the design of a mind map, which is intelligent, interactive, and capable of reducing the burden of teachers by aligning learning goals, supporting students' programming processes, and providing multimodal resources. We compared MindScratch to Scratch through a within-subject study involving 24 participants. The results show that MindScratch facilitates participants in achieving learning goals, enhancing code quality and creativity, and improving their CT skills. Additionally, we interviewed six programming educators to gather insights into the future of AI-driven educational tools. Our work offers insights and design considerations for building programming support tools to assist children with creative programming learning.